\documentclass{aa}
\usepackage[document]{ragged2e}
\usepackage{graphicx}
\usepackage{txfonts}
\usepackage{tabularx}
\usepackage{multirow}
\usepackage{amsmath}
\usepackage{mathtools}
\usepackage{amsfonts}
\usepackage{amssymb}

\begin{document}

\title{The negligible photodesorption of methanol ice and the active photon-induced desorption of its irradiation products.}
\author{G. A. Cruz-Diaz \inst{1,} \inst{2}, R. Mart\'\i{}n-Dom\'enech \inst{3}, G. M. Mu\~{n}oz Caro \inst{3}, and Y.-J. Chen \inst{4}}
\offprints{Gustavo A. Cruz-Diaz}
\institute{NASA Ames Research Center, Moffett Field, Mountain View, CA 94035, USA
\and Bay Area Environmental Research Institute, Petaluma, CA 94952, USA
\and Centro de Astrobiolog\'{\i}a, INTA-CSIC, Carretera de Ajalvir, km 4, Torrej\'on de Ardoz, 28850 Madrid, Spain
\and Department of Physics, National Central University, Jhongli City, Taoyuan County 32054, Taiwan \\
email: gustavo.a.cruzdiaz@nasa.gov; munozcg@cab.inta-csic.es} 
\date{Received - , 0000; Accepted - , 0000}
\authorrunning{G.A. Cruz-Diaz \& G.M. Mu\~{n}oz Caro}  
\titlerunning{Methanol non-photodesorption}

\abstract
%Context
{Methanol is a common component of interstellar and circumstellar ice mantles and is often used as an evolution indicator in star-forming regions. The observations of 
gas-phase methanol in the interiors of dense molecular clouds at temperatures as low as 10 K suggests that a non-thermal ice desorption must be active. Ice photodesorption was proposed 
to explain the abundances of gas-phase molecules toward the coldest regions.}
%Aims
{Laboratory experiments were performed to investigate the potential photodesorption of methanol toward the coldest regions.}
%Methods
{Solid methanol was deposited at 8 K and UV-irradiated at various temperatures starting from 8 K. The irradiation of the ice was monitored by means of infrared spectroscopy and the 
molecules in the gas phase were detected using quadrupole mass spectroscopy. Fully deuterated methanol was used for confirmation of the results.}
%Results
{The photodesorption of methanol to the gas phase was not observed in the mass spectra at different irradiation temperatures. We estimate an upper limit of 
3 $\times$ 10$^{-5}$ molecules per incident photon. On the other hand, photon-induced desorption of the main photoproducts was clearly observed.}
%Conclusions
{The negligible photodesorption of methanol could be explained by the ability of UV-photons in the 114 - 180 nm (10.87 - 6.88 eV) range to dissociate this molecule efficiently. 
Therefore, the presence of gas-phase methanol in the absence of thermal desorption remains unexplained. 
On the other hand, we find CH$_4$ to desorb from irradiated methanol ice, which was not found to desorb in the pure
CH$_4$ ice irradiation experiments.}

\maketitle

\justify 
\section{Introduction}

Methanol (CH$_3$OH) is abundant in ice mantles covering dust grains with relative-to-water abundances of 0.2-7 \% in comets, 5-12 \% in quiescent dense clouds, 1-30 \% around low mass 
protostars, and 5-30 \% around massive protostars (Mumma \& Charnley 2011; Boogert et al. 2015, and references therein). Gas-phase reactions that produce CH$_3$OH are very slow compared to grain 
surface chemistry for any physical conditions expected within protostellar systems. These reactions predict too low abundances (Geppert et al. 2006). Methanol is efficiently formed 
by irradiation of interstellar ice analog mixtures (Hudson \& Moore 1999, Watanabe et al. 2007). Alternatively, a proposed formation pathway on dust grains is through 
repeated hydrogenation of CO: 
\begin{equation}
\textrm{CO} \xrightarrow[\bf{H}]{} \textrm{HCO} \xrightarrow[\bf{H}]{} \textrm{H$_2$CO} \xrightarrow[\bf{H}]{} \frac{\textrm{CH$_2$OH}}{\textrm{CH$_3$O}} \xrightarrow[\bf{H}]{} \textrm{CH$_3$OH} \\
\end{equation}  
(Tielens \& Hagen 1982; Watanabe \& Kouchi 2002; Pontoppidan et al. 2004; Tielens 2005; Fuchs et al. 2009; Wirstr\"om et al 2011; Suutarinen et al. 2014). Wirstr\"om et al. (2011) concluded that the 
observed emission, which originates in extended cold envelopes of Young Stellar Objects (YSOs), is maintained by non-thermal desorption from dust grains. This means that non-thermal 
desorption processes 
like photodesorption, exothermic surface reactions, or cosmic-ray-induced heating need to be effective to explain the observed abundances of gas-phase methanol in dense and 
cold regions (Willacy \& Millar 1998; Shen et al. 2004; Garrod et al. 2007; \"Oberg et al. 2009; Leurini et al. 2010; Wirstr\"om et al 2011; Suutarinen et al. 2014). 

Leurini et al. (2010) 
reported a high concentration of gas-phase CH$_3$OH in the clumps of the Orion Bar compared with the inter-clump region. These clumps are cold, 10-20 K, and therefore thermal desorption 
is inhibited. They found that the column density of gas-phase methanol and formaldehyde in the clumps are of the same order of magnitude, while the column density of 
methanol in the inter-clump region is lower than that of formaldehyde. This could be the result of photodissociation of CH$_3$OH in the unshielded inter-clump gas. Guzm\'an et al. (2013) 
derived CH$_3$OH abundances, relative to the total number of hydrogen atoms, of $\sim$1.2 $\times$ 10$^{-10}$ and $\sim$2.3 $\times$ 10$^{-10}$ in the Horsehead PDR 
and its associated dense core, respectively. The H$_2$CO abundances in both positions displayed similar values ($\sim$2 $\times$ 10$^{-10}$). They observed that 
CH$_3$OH is depleted onto grains at the dense core. Therefore, CH$_3$OH is present in an envelope around the dense core position, while H$_2$CO is found in both the envelope and 
the dense core itself, in agreement with Leurini et al. (2010). They concluded that the two species are formed on the surface of dust grains and are subsequently photodesorbed 
into the gas-phase. Hence, photodesorption might be an efficient mechanism to release complex molecules in low-FUV-illuminated PDRs.

UV-photon-induced desorption of icy dust mantles is one of the non-thermal processes invoked in astrophysics to explain the molecular gas-phase abundances observed 
toward cold dark clouds and protoplanetary disks (Draine \& Salpeter 1979; Westley et al. 1995; Willacy \& Langer 2000; Andersson et al. 2005; Arasa et al. 2006; 
\"Oberg et al. 2007, 2009, 2010; Andersson \& van Dishoeck 2008; 
Mu\~noz Caro et al. 2010; Fayolle et al. 2011, 2013; Bahr \& Baragiola 2012; Bertin et al. 2012, 2013; Arasa et al. 2013, 2015; Yuan \& Yates 2013, 2014; Chen et al. 2014; 
Fillion et al. 2014; Mart\'\i{}n-Dom\'enech et al. 2015; van Hemert et al. 2015). 

UV-photon-induced desorption can take place through different mechanisms after absorption of the UV photons by the icy molecules. 
Absorption in the surface of the ice could lead to the direct desorption of the excited molecule itself, as proposed by van Hemert et al. (2015).
Alternatively, UV absorption in the layers right below the surface of the ice and subsequent energy transfer from the
excited molecule to neighboring molecules in the surface, allowing them to break the intermolecular bonds and leading
to their desorption, has been reported as an active photon-induced desorption mechanism (Öberg et al. 2007; Mu\~noz
Caro et al. 2010; Fayolle et al. 2011, 2013; Bertin et al. 2012, 2013; Fillion et al. 2014). This process was presented as 
indirect desorption induced by electronic transitions (DIET) in the literature, or ''kick-out`` desorption in van Hemert et al. (2015).

If the absorption of the UV photons leads to dissociation of the ice molecules, additional desorption pathways may play an important role, 
as suggested in theoretical works (e.g., Andersson \& van Dishoeck 2008, and Andersson et al. 2011). 
Hot atoms produced by photodissociation could transfer enough momentum to neighboring molecules, leading to a kick-out desorption.  
Photofragments formed with enough kinetic energy in the surface of the ice (due to the excess energy following photodissociation) can desorb directly after their formation. 
In addition, recombination of radicals can lead to the immediate desorption of the newly formed photoproduct, due to the excess energy of the recombinating 
fragments and/or the exothermicity of the recombination.  
Evidences of both processes have been experimentally found in Fayolle et al. (2013), Fillion et al. (2014), and Mart\'in-Dom\'enech et al. (2016).  
The key in these processes is that once a fragment or a molecule is formed on the surface as a consequence of the UV processing of the ice, 
it does not require energy from another photon to desorb. 
This can lead to a constant photon-induced desorption rate with photon fluence, as reported in Mart\'in-Dom\'enech et al. (2016), 
in contrast to the increasing desorption rate found for the products desorbing as a consequence of a previous UV absorption event in the subsurface region of the ice. 
These two processes are referred to as photon-induced chemical desorption or photochemidesorption in Mart\'in-Dom\'enech et al. (2015) and 
Mart\'in-Dom\'enech et al. (2016). 
Related processes were studied in Chakarov et al. (2001), where photon-induced desorption of photoproducts derived from UV-induced reactions were reported. 
If the photofragments or the photoproducts  
are formed in the subsurface, rather than on the surface, they may still induce desorption of 
nearby surface molecules \textit{via} energy transfer. 

\"Oberg et al. (2009) studied the ultraviolet irradiation of methanol. They focused in quantifying the UV-induced production rates of complex organic molecules like HCOOCH$_3$ and 
CH$_3$CH$_2$OH. In addition, they calculated a photodesorption yield of 2.1 $\pm$ 1.0 $\times$ 10$^{-3}$ molecules per incident photon based on their interpretation of the decrease of the 
IR band upon irradiation and they found a photodissociation cross section of 2.6 $\pm$ 0.9 $\times$ 10$^{-18}$ cm$^{2}$. They conclude that the methanol chemistry in ices is efficient 
enough to explain the observed abundances of complex organics around protostars.

This paper is structured as follows. The experimental protocol is described in Section 2. The experimental results regarding methanol ice photoprocessing are reported and discussed 
in Section 3. Special attention was payed to the detection of molecules desorbing from the ice, during irradiation or upon warm-up of the ice sample. Section 4 describes the two 
types of photon-induced desorption inferred from the experiments. The astrophysical implications and conclusions are reported in Section 5.

\section{Experimental protocol}
\label{Expe}

Laboratory experiments of solid methanol photoprocessing were conducted using the ISAC set-up. This set-up and the standard experimental protocol were described in Mu\~noz Caro et al. 
\cite{Caro1}. ISAC mainly consists of an 
ultra-high-vacuum (UHV) chamber, with pressure typically in the range P = 3-4.0 $\times$ 10$^{-11}$ mbar, where an ice layer made by deposition of a gas species onto a cold 
finger at 8 K, achieved by means of a closed-cycle helium cryostat, can be UV-irradiated. The evolution of the solid sample was monitored with \emph{in situ} transmittance 
FTIR spectroscopy and vacuum ultraviolet (VUV) spectroscopy. The chemical components used for the experiments described in this paper were: CH$_3$OH (liquid), Panreac Qu\'{\i}mica S.A. 99.9\%, and 
CD$_3$OD (liquid), Cambridge Isotope Laboratories, Inc (C.I.L.) 99.8\%. The deposited ice layer was photoprocessed with a microwave discharged hydrogen flow lamp (MDHL), 
from Opthos Instruments, with a hydrogen pressure of 0.40 $\pm$ 0.05 mbar. The Evenson cavity of the lamp is refrigerated with air. The source has a UV-flux of $\approx 2 \times 
10^{14}$ photons cm$^{-2}$ s$^{-1}$ at the sample position (Mu\~noz Caro et al. 2010). Routine measurements of the VUV-emission spectrum of the lamp were performed using a 
McPherson 0.2 m focal length VUV-monochromator (model 234/302). Based on our measurements, the mean energy of the MDHL was found to be 8.6 eV. The energy distribution and main bands of the UV-light produced 
by the MDHL are reported in Cruz-Diaz et al. (2014a). The characterization of the MDHL spectrum is discussed in more detail by Chen et al. (2014).

The methanol ice samples were irradiated with cumulative intervals that lead to total irradiation times of 5, 15, 30, 60, 120, and 240 min. Between each irradiation dose we performed 
FTIR spectroscopy to monitor the evolution of solid methanol as well as its photoproducts. As mentioned above, samples were deposited at 8 K and then irradiated or heated to the 
irradiation temperature of 30, 50, 70, 90, and 130 K. The deposited ice column density is shown in Table \ref{exp}. The ice column densities of methanol, methanol-D$_4$, and their 
photoproducts were calculated from their infrared absorption using the formula: $N = \frac{1}{\mathcal{A}} \int_{band} \tau_{\nu} d\nu$, where $N$ is the column density in molecules 
per cm$^{2}$, $\tau_{\nu}$ the optical depth of the band, $d\nu$ the wavenumber differential in cm$^{-1}$, and $\mathcal{A}$ the band strength in cm molecule$^{-1}$. The 
adopted band strengths were 1.8 $\times$ 10$^{-17}$ for the 1026 cm$^{-1}$ band of CH$_3$OH (d’Hendecourt \& Allamandola 1986), 7.0 $\times$ 10$^{-18}$ for the equivalent band of 
CD$_3$OD (Cruz-Diaz et al. 2014b), 6.2 $\times$ 10$^{-18}$ for the 1304 cm$^{-1}$ band of CH$_4$ (Pearl et al. 1991), 7 $\times$ 10$^{-18}$ for the 994 cm$^{-1}$ band of CD$_4$ 
(Kondo \& Saeki 1973), and 1.1 $\times$ 10$^{-17}$ for the 2139 cm$^{-1}$ band of CO (Jiang et al. 1975). The MDHL was always turned on to ensure a constant UV-flux and intensity. 
Quadrupole mass spectrometry (QMS) was used to detect the gas-phase molecules that desorbed thermally or photodesorbed from the ice during UV-irradiation. 

The integrated ion current measured by a QMS corresponding to a mass fragment $m/z$ of the molecules of a given species desorbed during ice irradiation experiments is 
proportional to the total number of molecules desorbed, and it can be calculated as follows:
\begin{equation}
A(m/z) = N(mol) \cdot k_{CO} \cdot \frac{\sigma^{+}(mol)}{\sigma^{+}(CO)} \cdot \frac{I_{F}(z)}{I_{F}(CO^+)} \cdot \frac{F_{F}(m)}{F_{F}(28)} \cdot \frac{S(m/z)}{S(28)}
\end{equation}
\label{correction}
where $A(m/z)$ is the integrated area below the QMS signal of a given mass fragment $m/z$ during photon-induced desorption in Ampers second (A s), $N(mol)$ the total number 
of desorbed molecules, in column density units ($cm^{-2}$), $k_{CO}$ is a proportionality constant ($7.9 \times10^{-24} \; \frac{A \; s}{molec. \; cm^{-2}}$ for our 
QMS in the case of CO molecules), $\sigma^+(mol)$ the ionization cross section for the first ionization of the species of interest with the incident electron energy of the mass 
spectrometer, $I_F(z)$ the ionization factor, i.e., the fraction of ionized molecules with charge $z$, $F_F(m)$ the fragmentation factor, i.e., the fraction of molecules of the 
isotopolog of interest leading to a fragment of mass $m$ in the mass spectrometer, and $S(m/z)$ the sensitivity of the QMS to the mass fragment $(m/z)$; see Mart\'\i{}n-Dom\'enech 
et al. (2015) for a detailed description of our QMS calibration method.

\begin{table}[ht!]
\centering
\caption{Methanol and deuterated methanol irradiation experiments performed for this study. One monolayer (ML) corresponds to 1 $\times$ 10$^{15}$ molecules cm$^{-2}$.}
\begin{tabular}{ccc}
\hline
\hline
CH$_3$OH&Irrad. temp.& Thickness\\
Sample No.&[K]&[ML]\\
\hline
1&8&124\\
2&30&155\\
3&50&128\\
4&70&131\\
5&90&117\\
6&130&100\\
\hline
CD$_3$OD&&\\
Sample No.&&\\
\hline
7&8&155\\
8&30&162\\
9&50&168\\
10&70&158\\
11&90&160\\
12&130&132\\
\hline
\end{tabular}
\label{exp}
\end{table}

\section{Experimental results and discussion}

\subsection{\textbf{CH$_3$OH/CD$_3$OD ice photoprocessing at 8 K}}

\begin{figure}[ht!]
\centering
\includegraphics[width=\columnwidth]{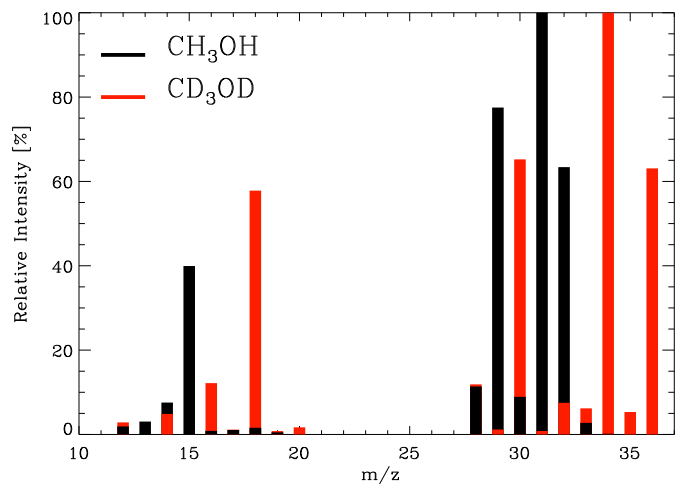}
\caption{Mass spectrum of methanol (black bars) and methanol-D$_4$ (red bars) obtained during deposition of the ice sample using our QMS at 8 K. The spectra are normalized 
relative to the main mass fragment (m/z = 31 for methanol and m/z = 34 for methanol-D$_4$).}
\label{MASAS}
\end{figure}

\begin{figure}[ht!]
\centering
\includegraphics[width=\columnwidth]{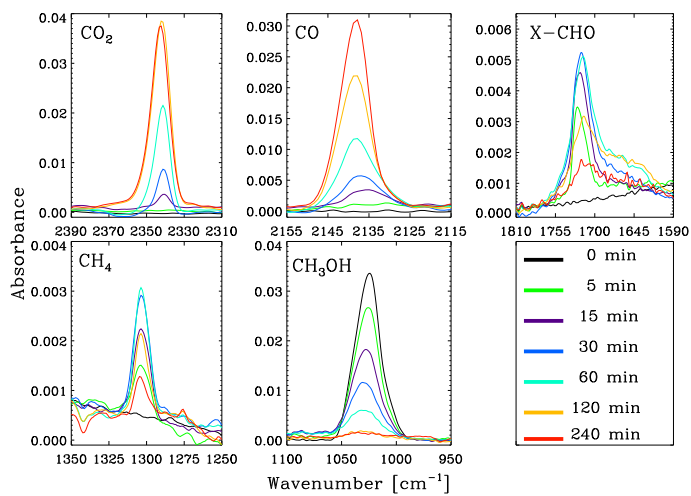}
\caption{Irradiation of pure CH$_3$OH ice at 8 K for different irradiation times. The panels show the decrease of the CH$_3$OH absorption and the formation of the main photoproducts.}
\label{IR-8K1}
\end{figure}

\begin{figure}[ht!]
\centering
\includegraphics[width=\columnwidth]{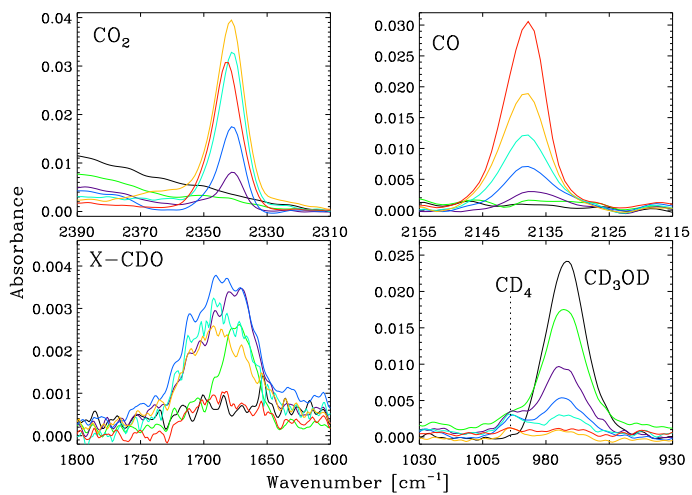}
\caption{Irradiation of pure CD$_3$OD ice at 8 K for the same irradiation times shown in Fig. \ref{IR-8K1}. The panels show the decrease of the CD$_3$OD absorption and the formation 
of the main photoproducts.}
\label{IR-8K2}
\end{figure}

\begin{figure}[ht!]
\centering
\includegraphics[width=\columnwidth]{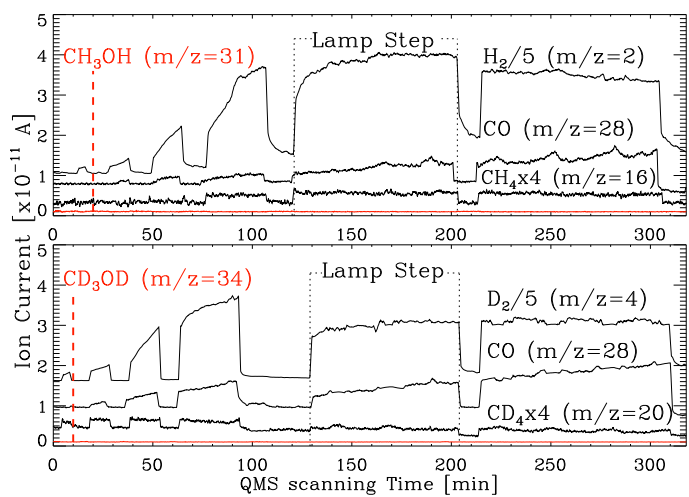}
\caption{Photon-induced desorption of the different photoproducts during UV-irradiation of methanol ice at 8 K. The different steps follow the turn on and turn off of the UV-lamp. 
The evolution of the photoproducts during irradiation is discussed in Section 3.2. Top: irradiation of pure CH$_3$OH ice. Bottom: irradiation of pure CD$_3$OD ice. The H$_2$ and D$_2$ 
signals were divided by 5 and the CH$_4$ and CD$_4$ signals were multiplied by 4 for clarity, to compare all the photoproducts. Small bumps in the curves are due to instabilities 
of the MDHL during irradiation, mainly because of the oscillations of the H$_2$ flow.}
\label{CH3OH-photo-des}
\end{figure}

Fig. \ref{MASAS} shows the mass spectra of methanol and fully deuterated methanol acquired with our QMS during the deposition of the samples. Methanol and methanol-D$_4$ were 
UV-irradiated first at 8 K. The bands of the observed photoproducts (CH$_4$/CD$_4$, CO, and CO$_2$) are displayed in Figs.~\ref{IR-8K1} and \ref{IR-8K2}. A broad band centered 
at 1720 cm$^{-1}$ is visible as a result of the contribution of different species. This band has been attributed to the X-CHO like molecules, e.g., H$_2$CO, HCOOH, 
CH$_3$CHO, and HCOOCH$_3$, see \"Oberg et al. (2009). The same band was observed for the deuterated species but shifted to 1680 cm$^{-1}$. During the TPD, the presence of HCOOH and 
CH$_3$CHO was confirmed based on their desorptions near 140 K of their main mass fragment, m/z 46 and m/z 43, respectively; but more dedicated experiments would be needed to study 
the thermal desorption of methanol photoproducts, similar to those reported by \"Oberg et al. (2009). Several species desorbing from the ice matrix during irradiation at 8 K were 
detected by QMS. We only estimated the number of molecules released to the gas for photon-induced desorbing 
species with the highest signal to noise ratio (H$_2$/D$_2$, CH$_4$/CD$_4$, and CO), see Fig.~\ref{CH3OH-photo-des}. 

Methanol (m/z = 31, the main mass fragment) photon-induced desorption was not observed, see top panel of Fig.~\ref{CH3OH-photo-des}, red trace. The methanol signal in 
the QMS does not show any perturbation when the MHDL is turned on or off, dotted lines. This is confirmed in the methanol-D$_4$ experiment (m/z = 34, the main mass fragment in 
Fig. \ref{MASAS}) red trace in bottom panel of Fig.~\ref{CH3OH-photo-des}. 

The number of molecules desorbing from the ice sample during irradiation was obtained using eq. 2. We performed irradiation experiments of pure CO ice to determine $k_{CO}$. 
Carbon monoxide was chosen because it cannot be directly dissociated by the UV-photons of our lamp, only about 5 \% of the UV-photons lead to rupture of the molecule via CO* + CO 
$\to$ CO$_2$ + C (Gerakines et al. 1996; Gerakines \& Moore 2001; Loeffler et al 2005; \"Oberg et al. 2007; Mu\~noz Caro et al. 2010; Chen et al. 2014). Therefore, the QMS signal of 
CO increases due to an almost pure photodesorption process. Photodesorption is the main responsible for both, the increase of the QMS signal of CO and the decrease of its IR 
absorption band. Then, the column density loss observed will be proportional to the signal area in the QMS during the irradiation and therefore the proportionality constant for CO, 
$k_{CO}$, can be estimated; see Mart\'\i{}n-Dom\'enech et al. (2015) for a detailed description of the process.

To relate the number of CO molecules photodesorbing from the ice sample with the number of molecules of any other photodesorbing species, some corrections are needed. 
We corrected the calculation using the electron-impact ionisation cross sections adapted from the National 
Institute of Standards and Technology (NIST Chemistry WebBook) at 70 eV for hydrogen, methane, and carbon monoxide: $\sigma_{H_2}$=1.021 \AA$^2$, $\sigma_{CH_4}$=3.524 \AA$^2$, 
and $\sigma_{CO}$=2.516 \AA$^2$. The electron-impact ionisation cross section of deuterium is similar to hydrogen, therefore we used the same value (1.021 \AA$^2$), 
see Rapp \& Englander-Golden (1965), tables 77 and 161 of Celiberto et al. (2001), and Yoon et al. (2009). We used the same value for methane and fully-deuterated methane as an 
approximation since $\sigma_{CD_4}$ was not found in the literature. The QMS filament has different detection sensitivity for each molecular mass. We calibrated for this effect 
using noble gases (He, Ne, and Ar) and extrapolated those values to obtain a mass-calibration curve for $S(m/z)$ of eq. 2 in the 4 to 40 amu range.

The QMS minimum detection signal is of the order of 3 $\pm$ 2 $\times$ 10$^{13}$ molecules cm$^{-2}$. This estimate was done using the CO signal in the QMS by applying 
the above procedure taking into account the detection time between two consecutive measurements, 5.89 s for our QMS. 
We divided the CO signal in the QMS by a constant to match the m/z = 31 signal, main fragment of CH$_3$OH in our QMS, in the same experiment. We found that: 
CO$_{signal}$ \textgreater 200 $\times$ CH$_3$OH$_{signal}$.  

Therefore, using the CO photodesorption yield measured in the same experiment (0.008 molecules per incident photon in Table 3), we calculated an upper limit for CH$_3$OH 
photodesorption of 3 $\times$ 10$^{-5}$ molecules per incident photon (after the QMS corrections mentioned in Section 2 were applied), i.e., about 2 orders of magnitude 
lower than \"Oberg et al. (2009) interpretation. It should be emphasized that this value is an upper limit for the photodesorption rate of methanol, which depends on the 
sensitivity of the QMS used in the experiment.

\begin{figure}[ht!]
\centering
\includegraphics[width=\columnwidth]{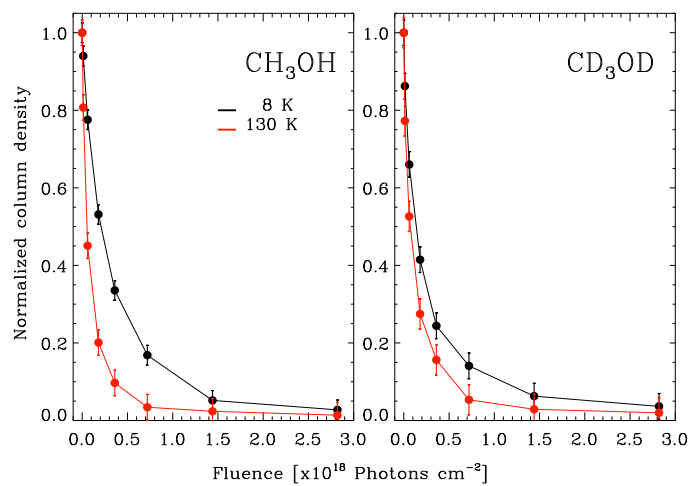}
\caption{Methanol and methanol-D$_4$ ice normalized column density as a function of fluence during the experiments 1 and 7 at 8 K (black trace) and experiments 6 and 12 at 130 K 
(red trace).}
\label{CH3OH-IR}
\end{figure}

Table \ref{yield} shows the photodissociated molecules calculated from the IR data, the absorbed photons calculated using the UV-absorption cross sections reported in 
Cruz-Diaz et al. (2014a) and (2014b), and the photodissociation yield for each irradiation interval. Using the equation for the photodissociation of species: 
\begin{equation}
N(t) = N(0) \exp(-\phi \cdot t \cdot \sigma_{des})
\end{equation}
(where N is the column density in cm$^{-2}$, $\phi$ is the flux of UV-photons in cm$^{-2}$s$^{-1}$, $t$ is the irradiation time, and $\sigma_{des}$ is the photodissociation 
cross-section by UV-photons), the methanol column densities plotted in Fig. \ref{CH3OH-IR}, and the UV-absorption cross sections (4.4 $\times$ 10$^{-18}$ cm$^2$ for CH$_3$OH 
and 4.6 $\times$ 10$^{-18}$ cm$^2$ for CD$_3$OD to calculate the amount of photons absorbed by the methanol ice) 
we found the methanol mean UV-dissociation cross sections to be 2.7 $\pm$ 0.7 $\times$ 10$^{-18}$ cm$^2$ for the above value of the incident photon flux; and 2.9 $\pm$ 1.9 $\times$ 
10$^{-17}$ cm$^2$ considering for this value only the absorbed photon flux. For methanol-D$_4$ we found the mean UV-dissociation cross sections to 
be 3.5 $\pm$ 1.2 $\times$ 10$^{-18}$ cm$^2$ based on the incident photon flux; and 2.4 $\pm$ 1.0 $\times$ 10$^{-17}$ cm$^2$ based on the absorbed photon flux.

\begin{table}[ht!]
\centering
\caption{Photodissociated molecules, absorbed photons, and photodissociation yield (in molecules per absorbed photon) at each irradiation interval for methanol and fully 
deuterated methanol at 8 K (experiments 1 and 7 in Table \ref{exp}).}
\tiny
\begin{tabular}{cccc}
\hline
\hline
Irrad.&Photodiss.&Absorbed&Photodiss.\\
Time&Molecules&Photons&Yield\\
min.&[$\times$ 10$^{16}$ molec./cm$^{2}$]&[$\times$ 10$^{16}$ phot./cm$^{2}$]&[molec./abs. phot.]\\
\hline
&&&\\
\multicolumn{4}{c}{CH$_3$OH}\\
&&&\\
5&2.8&5.3&0.5\\
10&3.0&10.4&0.3\\
15&2.4&16.1&0.2\\
30&2.1&32.8&0.06\\
60&1.5&67.5&0.02\\
120&0.3&136.1&0.002\\
\hline
&&&\\
\multicolumn{4}{c}{CD$_3$OD}\\
&&&\\
5&4.3&4.9&0.8\\
10&3.8&10.1&0.4\\
15&2.7&15.9&0.2\\
30&1.6&33.4&0.05\\
60&1.2&68.1&0.02\\
120&0.4&135.4&0.003\\
\hline
\end{tabular}
\label{yield}
\end{table}

\subsection{\textbf{CH$_3$OH/CD$_3$OD ice photoprocessing at various temperatures}}

\begin{figure}[ht!]
\centering
\includegraphics[width=\columnwidth]{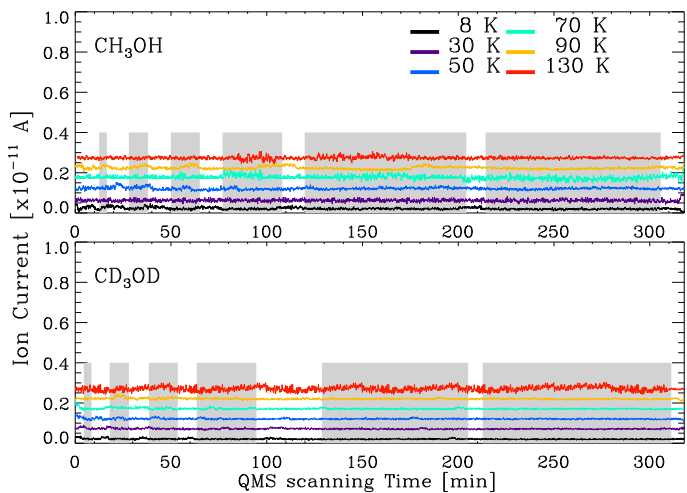}
\caption{QMS signals of CH$_3$OH ice (top) and CD$_3$OD ice (bottom) during the UV-irradiation, monitored with fragments the m/z 31 and 34, respectively. The color code indicates the 
different irradiation temperatures. The curves have been offset for clarity, since their signals were all at the same level except for the 130 K experiments. The gray areas 
represent UV-irradiation, when the lamp is turned on.}
\label{CH3OHCD3OD-irrad}
\end{figure}

\begin{table*}[ht!]
 \centering
 \caption{Desorbed products during the irradiation of methanol ice. Column \emph{[$\frac{mol.}{cm^2}$]} denotes the total number of desorbed molecules per square centimeter for 
 the total irradiation time, 240 minutes. Column \emph{[$\frac{mol.}{phot.}$]} denotes the average number of desorbed molecules per incident photon.}
 \tiny
 \begin{tabular}{cccccccc}
 \hline
 \hline
 &Irrad. Temp.&\multicolumn{2}{c}{H$_2$}&\multicolumn{2}{c}{CH$_4$}&\multicolumn{2}{c}{CO}\\
 &[K]&[$\frac{mol.}{cm^2}$]&[$\frac{mol.}{phot.}$]&[$\frac{mol.}{cm^2}$]&[$\frac{mol.}{phot.}$]&[$\frac{mol.}{cm^2}$]&[$\frac{mol.}{phot.}$]\\
 \hline
 \multirow{6}{*}{\rotatebox{90}{CH$_3$OH}}&&&&&&&\\
 &  8 &-$^*$&-$^*$&6.1 $\pm$ 0.9 $\times10^{15}$&0.002 &2.1 $\pm$ 0.6 $\times10^{16}$&0.008\\
 & 30 &-$^*$&-$^*$&5.8 $\pm$ 1.1 $\times10^{15}$&0.002 &5.2 $\pm$ 0.5 $\times10^{16}$&0.018\\
 & 50 &-$^*$&-$^*$&4.1 $\pm$ 1.3 $\times10^{16}$&0.014 &1.5 $\pm$ 0.3 $\times10^{17}$&0.056\\
 & 70 &-$^*$&-$^*$&2.9 $\pm$ 1.0 $\times10^{16}$&0.010 &6.6 $\pm$ 0.6 $\times10^{16}$&0.022\\
 & 90 &-$^*$&-$^*$&1.4 $\pm$ 0.3 $\times10^{16}$&0.005 &3.1 $\pm$ 0.6 $\times10^{16}$&0.011\\
 &130 &-$^*$&-$^*$&7.1 $\pm$ 0.4 $\times10^{15}$&0.002 &3.5 $\pm$ 0.4 $\times10^{15}$&0.0012\\
 \hline
 \hline
 &Irrad. Temp.&\multicolumn{2}{c}{D$_2$}&\multicolumn{2}{c}{CD$_4$}&\multicolumn{2}{c}{CO}\\
 &[K]&[$\frac{mol.}{cm^2}$]&[$\frac{mol.}{phot.}$]&[$\frac{mol.}{cm^2}$]&[$\frac{mol.}{phot.}$]&[$\frac{mol.}{cm^2}$]&[$\frac{mol.}{phot.}$]\\
 \hline
 \multirow{6}{*}{\rotatebox{90}{CD$_3$OD}}&&&&&&&\\
 &  8 &2.1 $\pm$ 0.7 $\times10^{18}$&0.7&4.5 $\pm$ 1.2 $\times10^{15}$&0.002 &2.4 $\pm$ 1.0 $\times10^{16}$&0.009\\
 & 30 &4.5 $\pm$ 0.4 $\times10^{18}$&1.6&4.3 $\pm$ 1.3 $\times10^{15}$&0.002 &5.6 $\pm$ 0.5 $\times10^{16}$&0.019\\
 & 50 &6.0 $\pm$ 0.6 $\times10^{18}$&2.1&2.6 $\pm$ 1.1 $\times10^{16}$&0.009 &9.5 $\pm$ 0.7 $\times10^{16}$&0.033\\
 & 70 &5.8 $\pm$ 0.3 $\times10^{18}$&2.0&1.3 $\pm$ 0.4 $\times10^{16}$&0.004 &3.2 $\pm$ 0.9 $\times10^{16}$&0.011\\
 & 90 &5.1 $\pm$ 0.2 $\times10^{18}$&1.8&9.9 $\pm$ 0.3 $\times10^{15}$&0.003 &9.2 $\pm$ 0.3 $\times10^{15}$&0.003\\
 &130 &4.4 $\pm$ 0.3 $\times10^{18}$&1.6&1.6 $\pm$ 0.4 $\times10^{16}$&0.005 &4.5 $\pm$ 1.0 $\times10^{15}$&0.0016\\
 \hline
 \end{tabular}
 \\
 (*) These values suffer from background H$_2$ contamination in the chamber. Therefore, only the values for D$_2$ in the deuteraed-methanol irradiation experiment are reliable.
 \label{tablePhoto}
 \end{table*}
 
The thermal desorption of pure CH$_3$OH ice occurs around 145 K in our ISAC setup, see Mart\'\i{}n-Dom\'enech et al. (2014). We found that the photodesorption of CH$_3$OH ice is 
negligible at 8 K. Photodissociation should be therefore the main effect at low temperatures. Fig.~\ref{CH3OH-IR} shows the decrease of the CH$_3$OH normalized column density as a 
function of UV-fluence at 8 K and 130 K. We observed that photodissociation of CH$_3$OH ice is more efficient at higher irradiation temperatures. 

The purpose of increasing the irradiation temperature was to study the role of temperature in the desorption. Desorption of methanol was also not observed during the 
UV-irradiation at temperatures higher than 8 K up to 130 K, see Fig. \ref{CH3OHCD3OD-irrad}. We increased the temperature at 130 K to observe if a change in the ice matrix structure 
(crystallization, see G\'alvez et al. 2009 and references therein) could induce a difference in the photodesorption process, but CH$_3$OH photodesorption was not observed at any 
temperature. 

Using the equation \ref{correction}, we were able to give an estimate of the total number of product molecules that desorbed from the methanol and methanol-D$_4$ samples, see Table 
\ref{tablePhoto}. Unlike the H$_2$ molecule, which desorption rate increases with temperature, in the case of CO and CH$_4$ photoproducts a maximum in the desorption is 
reached near 50 K. This behavior could, therefore, be attributed to the low sublimation temperatures of pure CO and CH$_4$ ices, respectively, $\sim$ 30 K and $\sim$ 40 K, allowing 
diffusion of a fraction of these volatiles in the methanol ice matrix, see Collings et al. (2004) and Mart\'\i{}n-Dom\'enech et al. (2014).

\subsubsection{\textbf{H$_2$/D$_2$ desorption}}

\begin{figure}[ht!]
\centering
\includegraphics[width=\columnwidth]{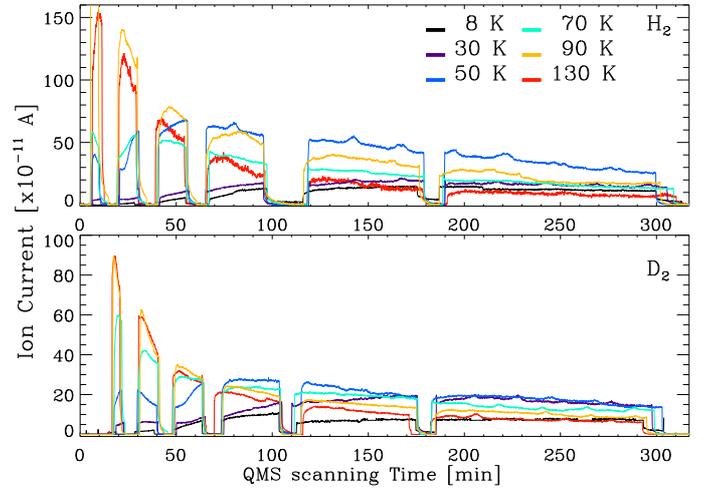}
\caption{Desorption of H$_2$ (top) and D$_2$ (bottom) during the irradiation of a pure methanol ice at different irradiation doses and temperatures indicated in the figure. Small 
bumps in the curves are due to instabilities of the MDHL during irradiation, mainly because of the oscillations of the H$_2$ flow.}
\label{H2D2-irrad}
\end{figure}

UV-irradiation of CH$_3$OH/CD$_3$OD produces H$_2$/D$_2$ as the main photoproduct. Hydrogen and deuterium were identified using the mass fragments m/z = 2 and m/z = 4, respectively. 
The experiments using deuterated methanol serve to discard the background H$_2$ contamination present in the chamber. Because of this H$_2$ contamination, reasonable values were 
only obtained monitoring D$_2$ desorption in the CD$_3$OD experiments. The desorption of H$_2$ is high due to the fact that most of the 
atomic hydrogen produced by photolysis (by the most likely process CH$_3$OH + h$\nu$ $\longrightarrow$ CH$_2$OH + H $\sim$ 4.26 eV) will diffuse through the ice and recombine with 
another hydrogen atom, then after reaching the surface it will desorb to the gas phase. The relative energies of H and D bonds are explained by differences in the zero-point 
vibrational energy. Raising the temperature tends to stabilize H over D bonds in methanol (Scheiner \& $\mathrm{\breve{C}}$uma 1996).  

Table \ref{tablePhoto} provides the total desorbed column density as a function of temperature for all the photoproducts of methanol ice. 
The desorption of D$_2$ is rather constant between 30 and 130 K. In the case of H$_2$, background contamination hinders a proper estimation of the desorption. 
Fig. \ref{H2D2-irrad} displays the desorption curves measured by QMS during the irradiation of CH$_3$OH and CD$_3$OD ice for different UV-doses and temperatures. 
A strong dependence with the irradiation temperature can be observed. The first 5 minutes of irradiation show the highest hydrogen desorption at 90 and 130 K, 
while the desorption at 8 K presents the weakest signal. Obviously, the higher the temperature, the easier the hydrogen molecules will diffuse, leading to an increase of the 
hydrogen QMS signal at the start of irradiation. 

\subsubsection{\textbf{CH$_4$/CD$_4$ desorption}}

\begin{figure}[ht!]
\centering
\includegraphics[width=\columnwidth]{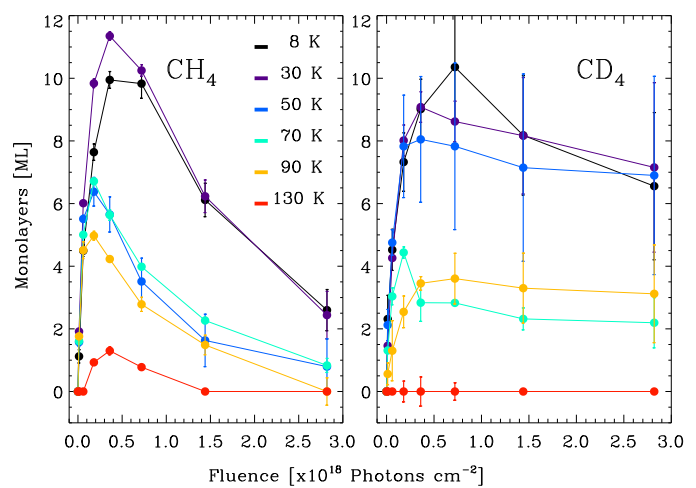}
\caption{Column density of CH$_4$/CD$_4$ as a function of UV-fluence during the irradiation of CH$_3$OH/CD$_3$OD ice measured by IR spectroscopy.}
\label{CH4CD4-IR}
\end{figure}

FTIR spectroscopy served to measure the column density of produced CH$_4$/CD$_4$ during irradiation at different temperatures, see Fig. \ref{CH4CD4-IR}. At 8 K, the 
CH$_4$ column density reaches a maximum after 30 minutes of irradiation that corresponds to about 8 \% of the deposited methanol. However, methane desorption is 
observed from the start of irradiation. This effect is better observed during the irradiation of CD$_3$OD, see Fig. \ref{CH3OH-photo-des} bottom panel. 
After a fluence of 3.6 $\times$ 10$^{17}$ photon cm$^{-2}$ (30 min. irradiation) the abundance of H-bearing molecules like CH$_4$ and X-CHO starts to decrease, see Fig. \ref{IR-8K1}. 
The decrease in the column density of CH$_4$ can be attributed to photodissociation as well as photon-induced desorption.
Table \ref{tablePhoto} reports the desorbed methane and methane-D$_4$ column density at 8 K. We measured a mean photon-induced desorption yield of 0.002 molecules per incident 
photon. We calculated the formation cross-section $\sigma_{form}$ of methane using the equation:
\begin{equation}
\frac{dN}{dt} = \phi \cdot N_p \cdot \sigma_{form} 
\end{equation}
where $N$ is the column density of the photoproduct (methane, in cm$^{-2}$), $N_p$ is the column density of methanol in cm$^{-2}$, and $\phi$ is the flux of UV-photons in 
photons cm$^{-2}$ s$^{-1}$. 
We found the mean formation cross sections of CH$_4$ to be 3.1 $\pm$ 1 $\times$ 10$^{-21}$ cm$^2$ for the incident photon flux and 3.5 $\pm$ 0.7 $\times$ 10$^{-19}$ cm$^2$ 
for the absorbed photon flux by methanol ice. For CD$_4$ the mean formation cross sections are 3.0 $\pm$ 1.5 $\times$ 10$^{-21}$ cm$^2$
for the incident photon flux and 4.0 $\pm$ 1.2 $\times$ 10$^{-19}$ cm$^2$ for the absorbed photon flux by methanol-D$_4$. The methane production is 
temperature dependent and it reaches a maximum between 15 and 30 minutes of irradiation. In the case of methane-D$_4$, it seems to display the same behavior, but due to the overlap 
with the methanol-D$_4$ band, the column density measurements become less accurate, see Fig. \ref{IR-8K2}.

\begin{figure}[ht!]
\centering
\includegraphics[width=\columnwidth]{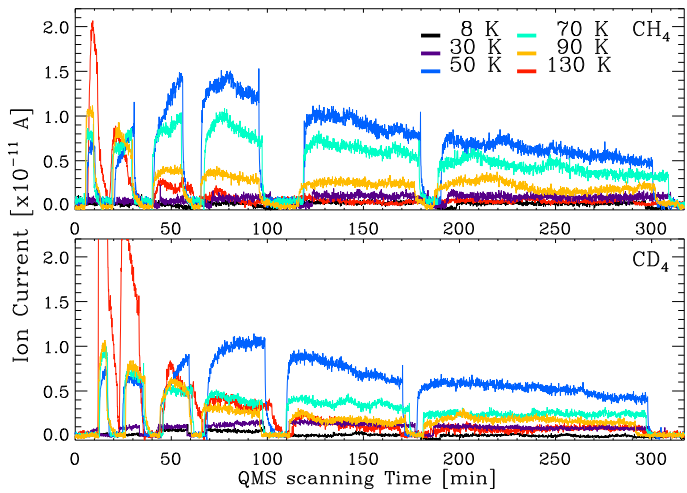}
\caption{Desorption of CH$_4$ (top) and CD$_4$ (bottom) during the irradiation of a pure CH$_3$OH/CD$_3$OD ice at different irradiation doses and temperatures indicated in the figure. 
Small bumps in the curves are due to instabilities of the MDHL during irradiation, mainly because of the oscillations of the H$_2$ flow.}
\label{CH4CD4-irrad}
\end{figure}

The desorption of CH$_4$/CD$_4$ during irradiation of CH$_3$OH/CD$_3$OD ice is shown in Fig. \ref{CH4CD4-irrad}. CH$_4$ and CD$_4$ were monitored using the mass fragments 
m/z = 16 and m/z = 20. Contamination due to other species sharing the same mass fragment like atomic oxygen, m/z = 16, and deuterated water, m/z = 20, was negligible. This was found 
tracing the mass fractions m/z = 32 (O$_2$) and m/z = 18 (OD). The desorption of CH$_4$/CD$_4$ is temperature dependent; at 130 K the registered QMS signal is much higher than at 
lower temperatures. This behavior is contrary to the production of methane in Fig. \ref{CH4CD4-IR}. Diffusion will increase with temperature, which reduces the amount of methane 
retained inside the methanol ice matrix. At 8 K and 30 K, photon-induced desorption is responsible for the small increase in the CH$_4$ signal. At 50 K methane reaches its maximum 
desorption because thermal desorption is active, see CH$_4$ and CD$_4$ columns from Table \ref{tablePhoto}. Above the CH$_4$ sublimation temperature, its QMS signal presents a broad 
bump that can be appreciated from the begining to the end of the total irradiation time, while UV-irradiation produces a step function-like feature that is clearly observed at 
8 K (Fig. \ref{CH3OH-photo-des}) and 30 K (Fig. \ref{CH4CD4-irrad}).

The likely process leading to desorption of CH$_4$ starts with the photodissociation of CH$_3$OH giving CH$_3$ + OH, a process that requires 92.1 kcal mol$^{-1}$ (4.0 eV). 
Other dissociation routes expected to occur have 
a lower probability because the energies required are higher: 96.1 kcal mol$^{-1}$ for CH$_2$OH + H and 104.6 kcal mol$^{-1}$ for CH$_3$O + H (Blanksby \& Ellison 2003). The fast 
exothermic reaction of this methyl radical with hydrogen on the ice surface, the later generated from photodissociation of CH$_3$OH into CH$_2$OH/CH$_3$O + H, 
leads to methane formation: CH$_3$ + H $\to$ CH$_4$ + 2.23 eV. This excess energy corresponds to the 
enthalpy of the reaction using the standard enthalpies of formation of methane (-0.69 eV) and the methyl radical (1.54 eV), see Chase (1998). We note that this excess energy 
of 0.69 eV is larger than the binding energy of methane molecules in a methanol dominated ice matrix, since this value will be no higher than the binding energy of pure methanol ice 
of about 4355 K (or 0.38 eV), see e.g., Mart\'\i{}n-Dom\'enech et al. (2014). 

\subsubsection{\textbf{CO desorption}}

\begin{figure}[ht!]
\centering
\includegraphics[width=\columnwidth]{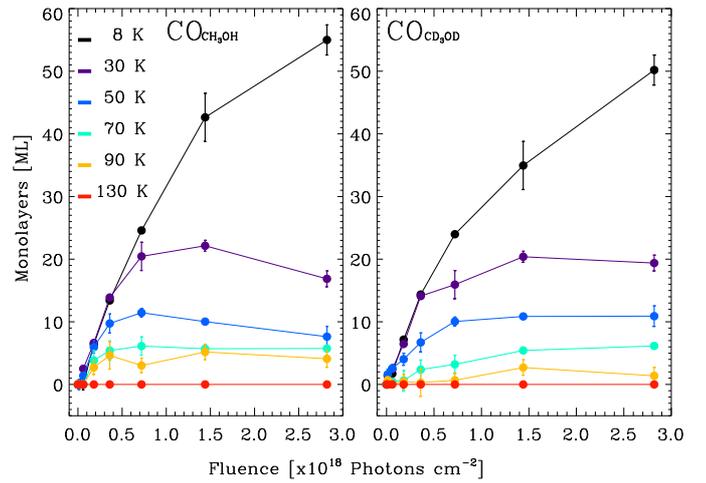}
\caption{Column density of CO as a function of UV-fluence during the irradiation of CH$_3$OH/CD$_3$OD ice measured by IR spectroscopy.}
\label{CO-IR}
\end{figure}

\begin{figure}[ht!]
\centering
\includegraphics[width=\columnwidth]{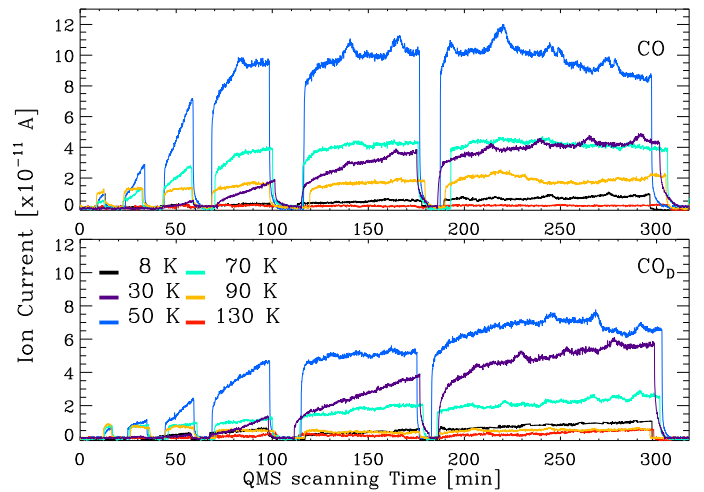}
\caption{Desorption of CO during the irradiation of pure CH$_3$OH (top) and CD$_3$OD (bottom) ice at different irradiation doses and temperatures indicated in the figure. Small bumps 
in the curves are due to instabilities of the MDHL during irradiation, mainly because of the oscillations of the H$_2$ flow.}
\label{CO-irrad}
\end{figure}

Fig. \ref{CO-IR} represents the column density of CO as a function of UV-fluence for the irradiation of pure methanol and methanol-D$_4$ ice. It can be observed that both 
experiments behave rather similarly. At 8 K, the CO column density reaches a maximum that corresponds to about 44 \% of the deposited methanol for 4 hours of irradiation time. This 
means that almost half of the methanol dissociates to produce CO by dehydrogenation in a sequence that requires 4 photons (CH$_3$OH $\xrightarrow{UV}$ CH$_2$OH $\xrightarrow{UV}$ 
H$_2$CO $\xrightarrow{UV}$ HCO $\xrightarrow{UV}$ CO), see \"Oberg et al (2009). The desorption of 
carbon monoxide is observable after a fluence of 6 $\times$ 10$^{16}$ photons cm$^{-2}$ (5 minutes) at 8 K. Using Eq. 4, we found that the mean formation cross sections of CO in 
irradiated methanol was 3.0 $\pm$ 2.0 $\times$ 10$^{-21}$ cm$^2$ considering the 
incident photon flux and 8.3 $\pm$ 2.2 $\times$ 10$^{-19}$ cm$^2$ considering the absorbed photon flux by methanol. For CO produced by CD$_3$OD 
irradiation the mean formation cross sections are 3.1 $\pm$ 1.1 $\times$ 10$^{-21}$ cm$^2$ considering the incident photon flux and 7.7 $\pm$ 2.0 $\times$ 10$^{-19}$ cm$^2$ 
considering the absorbed photon flux by methanol-D$_4$ ice. The column density of carbon monoxide present in the ice decreases drastically at 30 K, since the 
thermal desorption of CO occurs near 28 K in our set-up, the maximum is achieved at 2 hours of irradiation with an abundance that corresponds to about 14 \% of the deposited 
methanol. The desorption of CO is shown in Fig. \ref{CO-irrad}.

\section{Two patterns in the photon-induced desorption of photoproducts}

The irradiation of ice at temperatures near or above the sublimation temperature of the photoproducts leads to the release of ice molecules to the gas, mainly caused by thermal 
desorption, while radiative processes may also contribute to this desorption to some extent. Below, we will mainly discuss the desorption observed at the lowest temperatures, in 
particular at 8 K, corresponding to the coldest regions in dark clouds where thermal desorption is inhibited. 

\begin{figure}[ht!]
\centering
\includegraphics[width=\columnwidth]{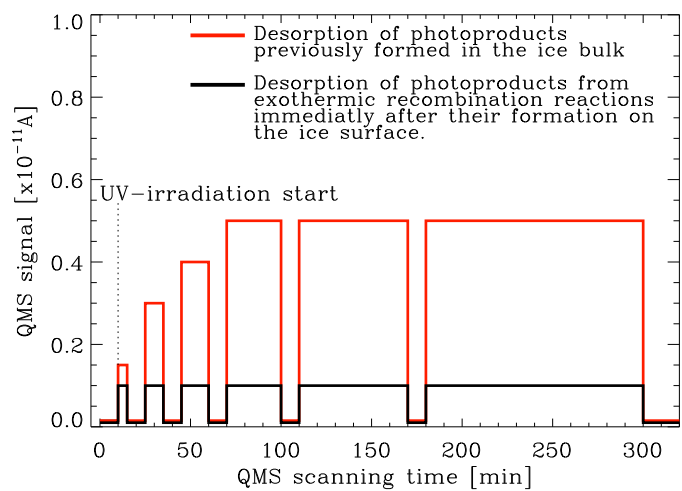}
\caption{QMS signal representation of two photon-induced desorption processes taking place in the UV-irradiation of an ice mantle.
Red trace: desorption of photoproducts previously formed in the ice bulk. 
Black trace: desorption of photofragments or photoproducts from exothermic recombination reactions immediatly after their formation on the ice 
surface.}
\label{QMS-F-Q}
\end{figure}

The pure methanol ice irradiation experiments led to the formation of several species, which were observed by IR spectroscopy. 
Some products were found to desorb into the gas phase during UV irradiation of the ice samples, leading to an increase in the ion current 
of the corresponding m/z fragment. 
Two different patterns in the behavior of the ion currents can be recognized in Fig. \ref{CH3OH-photo-des}, which are outlined in Fig. \ref{QMS-F-Q}. 
These two different patterns should correspond to different photon-induced desorption mechanisms operating in the ice (see Mart\'in Dom\'enech et al. 2015, 2016). 

Most of the photoproducts show an increase of their ion currents with fluence. 
As stated in Mart\'in Dom\'enech et al. (2015, 2016), this is tentatively associated to the desorption of molecules previously formed 
and accumulated in the ice bulk, that became exposed to the surface after the ice monolayers on top were removed upon continued irradiation. 
The subsequent absorption of a UV photon by a nearby molecule, 
followed by the energy transfer to the surface molecules 
could lead to their photon-induced desorption. 
Therefore, this mechanism for photon-induced desorption requires two or more photons: at least one photon to form the product molecule in the ice bulk, 
 and another photon to break the intermolecular bonds leading to its desorption once the 
molecule becomes eventually exposed to the ice surface. 
This pattern of photon-induced desorption 
becomes observable when the product concentration stored in the ice reaches a minimum value,
as observed by QMS. 
A good example is the desorption of the CO photoproduct  in irradiated methanol ice,
see Fig. \ref{CH3OH-photo-des}. 
 
A substantially different pattern of desorption was observed in Fig. \ref{CH3OH-photo-des} for CH$_{4}$ and CD$_{4}$. 
In this case, during the various intervals of irradiation the ion current of the photoproducts led to a step function: 
there is a sharp rise of the QMS signal at the beginning of the irradiation and a sharp drop of the signal when irradiation is stopped 
(step function). 
The value of the QMS signal tends to remain constant during irradiation, meaning that the desorption rate is also constant. 
Since this mechanism is not an accumulative process, 
photon-induced desorption is probably taking place immediatly after the recombination of photofragments resulting from  
photodissociation of methanol molecules in the ice surface, as proposed in previous sections. 
This alternative desorption mechanism, along with the desorption of excited photofragments right after their formation, 
was refered to as photochemidesorption in Mart\'\i{}n-Dom\'enech et al. (2015, 2016). 
In this case, a photoproduct formed in the ice surface directly desorbs because the excess in kinetic energy overcomes the binding energy of the ice. 
Since this occurs only when the molecule is formed on the ice surface, the expected rate of desorption is low. 
Therefore, it may be difficult to detect depending on the 
sensibility of the QMS, vacuum conditions, etc. 

The latter non-thermal desorption process may enable the release to the gas phase in cold regions of certain molecules that are not able to desorb 
via energy transfer from nearby photo-excited molecules.
This provides an alternative route to account for the presence of methane in cold interstellar regions, 
provided that this molecule is formed by irradiation of a different precursor species in the ice,
since  
irradiation of pure methane ice experiments performed at 8 K did not lead to an observable photodesorption of this molecule. 

Table \ref{Photochem} shows the main difference between these two photon-induced desorption processes, which is also noticiable through the 
photon-induced desorption rate of the different molecules, as explained in Sect. 1. 
In this case, the CO desorption increases with 
irradiation time while CH$_4$ desorption remains rather stable.     

\begin{table}[ht!]
\centering
\caption{Desorption yields for CD$_4$ and CO at 8 K during pure deuterated methanol ice irradiation.}
\tiny
\begin{tabular}{ccc}
\hline
\hline
Irrad. time&CD$_4$&CO\\
&&\\
min.&\multicolumn{2}{c}{[molec./phot.]}\\
\hline
&&\\
5  &0.001&0\\
10 &0.002&0.002\\
15 &0.002&0.005\\
30 &0.003&0.007\\
60 &0.002&0.008\\
120&0.002&0.009\\
\hline
\end{tabular}
\label{Photochem}
\end{table}

\section{Astrophysical implications and conclusions}
\label{Astro}

CH$_{3}$OH ice probably forms in dense cloud cores; it is absent from the cloud edges, but it is often abundant toward protostars. Because of its formation deep into the cloud, 
methanol ice mantles are shielded from the external UV-field during most of their lifetime (Charnley et al. 1992; Garrod et al. 2008). We found no observable photodesorption 
of CH$_{3}$OH or CD$_3$OD. Using a QMS for the detection of molecules in the gas phase, the estimated upper limit for the methanol photodesorption is 3 $\times$ 10$^{-5}$
molecule per incident photon. This upper limit is about 2 orders of magnitude lower that the value provided by \"Oberg et al. (2009), based on their interpretation of the 
decrease of the IR band 
upon irradiation. However, the two UV-dissociation cross section values for CH$_3$OH ice found by \"Oberg et al. (2009) and this work are very similar: 2.6 ($\pm$0.9) 
$\times$ 10$^{-18}$ cm$^2$ and 2.7 ($\pm$0.7) $\times$ 10$^{-18}$ cm$^2$ based on the incident photon flux, respectively.

Our work using a combination of FTIR and QMS shows that CH$_{3}$OH ice has a negligible photodesorption, since it was not detected by the QMS, which has a higher 
sensitivity than the IR technique used by \"Oberg et al. (2009) to estimate the methanol photodesorption rate in an indirect way. We consider that the value provided by \"Oberg et 
al. (2009) should be regarded as an upper limit. Here we provide a more stringent value of this upper limit for CH$_3$OH photodesorption. Therefore, a different non-thermal desorption 
process (or a combination of processes) is required to explain the amount of methanol detected in the gas phase toward dense interstellar clouds: this could be desorption due to 
release of chemical heat following recombination of two thermalized radicals or cosmic-ray induced desorption which could be simulated experimentally. Even for a total 44 \% 
conversion of CH$_3$OH to CO, a CH$_3$OH photodesorption route due to indirect desorption through CO excitation was not observed, see Bertin et al. (2013) and Fillion et al. (2014) 
for a detailed description of the process. The efficient photodissociation of CH$_3$OH leads to the formation and desorption of various products but inhibits the 
photodesorption of CH$_3$OH molecules.

Although direct photodesorption of CH$_{3}$OH/CD$_3$OD was not observed, the H$_2$/D$_2$, CO, and CH$_4$/CD$_4$ photoproducts were found to desorb by two different mechanisms, 
leading to different patterns in the observed photon-induced desorption yields. Since non-thermal desorption in dense clouds competes with accretion of molecules, photoproducts  
formed in the bulk of interstellar ice mantles may never be exposed to the ice surface. Therefore, the influence of the different mechanisms in the desorption yields could be 
limited to some extent. In any case, UV-irradiation of ice mantles 
can therefore contribute to recreate their gas phase abundances in cold inter- and circumstellar regions, being molecular hydrogen the main photoproduct in the case of 
methanol ice. Since the irradiation of pure CH$_4$ ice did not lead to an observable photodesorption of this molecule, the photochemidesorption of CH$_4$ by CH$_3$OH ice irradiation 
provides an alternative route to release methane to the gas in cold regions.

Since submission of our manuscript in June, 2015, a paper was published on methanol irradiation (see Bertin et al. 2016), although both works are not directly comparable since 
we used a continuum radiation source and they employed monochromatic synchrotron light with very low doses.

\begin{acknowledgements}
This research was financed by the Spanish MINECO under project AYA2011-29375, AYA2014-60585-P, and CONSOLIDER grant CSD2009-00038. This work was partially supported by NSC 
grants NSC99-2112-M-008-011-MY3 and NSC99-2923-M-008-011-MY3, and the NSF Planetary Astronomy Program under Grant AST-1108898.
\end{acknowledgements}

\end{document}